\documentclass[prd,a4paper,twocolumn,preprintnumbers,superscriptaddress,nofootinbib]{revtex4}
\pdfoutput=1
\usepackage{amsmath}
\usepackage{graphicx}
\usepackage{xspace}
\usepackage{color}
\usepackage{units}
\usepackage{slashed}
\usepackage[hyperfootnotes=false]{hyperref}

\def \L {\mathcal{L}} 

\def \epsilon {\varepsilon} 

\newcommand{\hc}{\ensuremath{\text{h.c.}}}

\newcommand{\dd}{\mathrm{d}}

\allowdisplaybreaks

\makeatletter
\newcommand\footnoteref[1]{\protected@xdef\@thefnmark{\ref{#1}}\@footnotemark}
\makeatother

\begin{document}

\title{Leptogenesis and neutral gauge bosons}

\preprint{ULB-TH/16-15}

\author{Julian Heeck}
\email{Julian.Heeck@ulb.ac.be}
\affiliation{Service de Physique Th\'eorique, Universit\'e Libre de Bruxelles, Boulevard du Triomphe, CP225, 1050 Brussels, Belgium}

\author{Daniele Teresi}
\email{Daniele.Teresi@ulb.ac.be}
\affiliation{Service de Physique Th\'eorique, Universit\'e Libre de Bruxelles, Boulevard du Triomphe, CP225, 1050 Brussels, Belgium}

\hypersetup{
    pdftitle={
		Leptogenesis and neutral gauge bosons},
    pdfauthor={Julian Heeck, Daniele Teresi}
}


\begin{abstract}
We consider low-scale leptogenesis via right-handed neutrinos $N$ coupled to a $Z'$ boson, with gauged $U(1)_{B-L}$ as a simple realization. Keeping the neutrinos sufficiently out of equilibrium puts strong bounds on the $Z'$ coupling strength and mass, our focus being on light $Z'$ and $N$, testable in the near future by SHiP, HPS, Belle II, and at the LHC. 
We show that leptogenesis could be robustly falsified in a large region of parameter space by the double observation of $Z'$ and $N$, e.g.~in the channel $pp\to Z' \to NN$ with displaced $N$-decay vertex, and by several experiments searching for light $Z'$, according to the mass of $N$.
\end{abstract}

\maketitle


\section{Introduction}

Right-handed neutrinos $N_R$ are a popular minimal extension of the Standard Model (SM) that can generate neutrino masses, the baryon asymmetry of the Universe, and even dark matter~\cite{Canetti:2012kh}. While neutrino masses via the seesaw mechanism~\cite{Minkowski:1977sc} and baryogenesis via leptogenesis~\cite{Fukugita:1986hr} are typically assumed to arise from right-handed neutrino masses above $\unit[10^{8}]{GeV}$~\cite{Davidson:2002qv,Hambye:2003rt}, low-scale realizations such as resonant leptogenesis~\cite{Pilaftsis:1997jf,Pilaftsis:2003gt,Dev:2014laa} exist and provide experimental testability.
It was recently shown in Ref.~\cite{Hambye:2016sby} that an absolute lower bound $M_N > \unit[2]{GeV}$ exists, assuming a thermal population of $N_R$ in the early Universe, while a non-thermal population can lead to successful leptogenesis even for lower $M_N$.

If the right-handed neutrinos $N_R$ are part of a more complete model, they are often endowed with additional interactions, potentially threatening the generation of a lepton asymmetry (for example Sakharov's out-of-equilibrium condition). A prime example here are left--right symmetric models based on the gauge group $SU(2)_L\times SU(2)_R\times U(1)_{B-L}$~\cite{Mohapatra:1979ia,Mohapatra:1980yp}, which couple the $N_R$ to a new charged gauge boson $W_R^-$. Successful leptogenesis then requires $M_{W_R} > \unit[10]{TeV}$; otherwise the gauge interactions with strength $g_R = g_L = e/\sin\theta_W$ heavily dilute the asymmetry~\cite{Carlier:1999ac,Frere:2008ct,Dev:2015vra}.
Interactions via a \emph{neutral} gauge boson $Z'$ are typically considered less dangerous because all such processes require \emph{two} $N_R$ instead of just one~\cite{Cosme:2004xs,Frere:2008ct}. A well-motivated simple model for interactions of this sort can be obtained by promoting the anomaly-free global symmetry $U(1)_{B-L}$ of the SM to a gauge symmetry, which actually \emph{requires} three right-handed neutrinos for consistency.
Previous work on leptogenesis in a $U(1)_{B-L}$ context has been performed in Refs.~\cite{Plumacher:1996kc,Frere:2008ct,Racker:2008hp,Blanchet:2009bu,Iso:2010mv}.

In this article we (re)evaluate the constraints on the $U(1)_{B-L}$ parameter space coming from successful leptogenesis and focus in particular on the case of light $N_R$ and $Z'$, potentially testable and falsifiable in the future.

\section{Framework}
\label{sec:framework}

We consider the simplest SM extension that has a $Z'$ boson coupled to (three) right-handed neutrinos $N_R$, based on the anomaly-free gauge group $U(1)_{B-L}$.\footnote{Other realizations of the relevant coupling $Z_\mu' \overline{N}\gamma^\mu \gamma_5 N$ can be found for example in left--right models with $M_{W_R} \gg M_{Z_R}$ or in models with gauged $U(1)_L$ and would not change our conclusions much.} To provide Majorana masses to $N_R$ we add one SM-singlet complex scalar $\Phi$ with $B-L$ charge $+2$:
\begin{align}
\begin{split}
\L &= i \overline{N}_R \slashed{D} N_R + |D_\mu \Phi|^2 - \lambda_S \left(\frac{w^2}{2}-|\Phi|^2\right)^2\\
&\quad -\left( y \overline{L} H N_R + \lambda_N \overline{N}_R^c N_R \Phi + \hc \right) ,
\end{split}
\end{align}
suppressing flavor indices. Without loss of generality we can assume $\lambda_N$ to be diagonal with positive entries.
In unitary gauge, $\Phi = (w+S)/\sqrt{2}$, one physical scalar $S$ with mass $M_S = \sqrt{2\lambda_S} w$ remains, while the vacuum expectation value (VEV) $w$ induces a $Z'$ mass $M_{Z'} = 2 g' w$ and a mass matrix $\mathcal{M}_N =  \sqrt{2}\lambda_N w$ for the sterile Majorana neutrinos $N=N_R + N_R^c$.
After electroweak symmetry breaking the right-handed neutrinos mix with the left-handed ones and provide light neutrino masses in the standard way.\footnote{Active--sterile neutrino mixing $\theta$ will lead to interaction terms $Z' N \nu$ that are \emph{linear} in $N$ and thus potentially more dangerous for leptogenesis in direct analogy to the $W_R^-$ scenario. Since these couplings are however suppressed by $g' \theta$ they turn out to be subleading.} 
Note that we further ignore mixing of $S$ with the SM Brout--Englert--Higgs boson as well as kinetic mixing of the $Z'$, which is a conservative choice.

\section{Leptogenesis}
\label{sec:leptogenesis}

\subsection{Boltzmann equations}\label{sec:boltz}

Assuming a Universe that started with total $B-L$ charge zero -- possibly after an inflationary period -- a non-zero baryon asymmetry can be 
dynamically generated at temperatures below the $B-L$ thermal phase transition, which typically occurs at a temperature of the order of  the $B-L$ breaking VEV $w$,\footnote{In the presence of additional scalar fields the critical temperature can be much larger than the VEV; see e.g.~Ref.~\cite{Pilaftsis:2008qt}.}
and above the sphaleron decoupling temperature $T_\text{sph} = \unit[131.7]{GeV}$~\cite{D'Onofrio:2014kta}. An (unflavored) $B-L$ asymmetry requires not only $w\neq 0$, but also $y\neq 0 \neq \lambda_N$, CP-violating phases in $y$, as well as out-of-equilibrium dynamics for $N$.
While these conditions are most commonly satisfied by considering a slow decay $N\to L H$ compared to the expanding Universe, they can also be satisfied via the decay $H\to L N$~\cite{Hambye:2016sby} thanks to thermal effects, making it possible to consider $N$ masses as low as $\unit[2]{GeV}$.
Our analysis follows Ref.~\cite{Hambye:2016sby}, but includes the new interactions involving $Z'$ and $S$, 
\begin{align}
NN \leftrightarrow {Z'}^{(\ast)} \leftrightarrow f\bar f \,, &&
NN \leftrightarrow Z' Z'\,,\, Z' S\,,\, SS\,,
\end{align}
which potentially keep $N$ in thermal equilibrium with the SM and hence dilute the resulting lepton asymmetry. 
We collect the cross sections in App.~\ref{sec:formulae}. Since both $Z'$ and $S$ are in thermal equilibrium with the SM in the regime considered in this work, we only have to consider the Boltzmann equations for the $N$ number density $n_N$ and the total lepton asymmetry $n_L$, given respectively by
\begin{align}\label{eq:boltzmann}
&\frac{n_\gamma H_N}{z} \, \frac{\dd \eta_N}{\dd z} \ = \ -\; \bigg[\bigg(\frac{\eta^N}{\eta_N^{\rm eq}}\bigg)^2  - 1 \bigg] \, 2 \, \gamma_{NN} \nonumber\\
&- \bigg(\frac{\eta^N}{\eta_N^{\rm eq}} \, - \, 1 \bigg)\Big[ \gamma_D + 2 (\gamma_{Hs}+\gamma_{As})  + 4 (\gamma_{Ht} + \gamma_{At})\Big] \;,\\
&\frac{n_\gamma H_N}{z} \, \frac{\dd \eta_L}{\dd z} \ = \ \gamma_D \bigg[\bigg(\frac{\eta^N}{\eta_N^{\rm eq}}-1 \bigg) \epsilon_{\rm CP}(z) - \frac{2}{3} \eta_L \bigg] \nonumber\\
& \qquad\quad -\frac{4}{3} \eta_L \bigg[2 (\gamma_{Ht}+\gamma_{At}) + \frac{\eta^N}{\eta_N^{\rm eq}} (\gamma_{Hs} + \gamma_{As})\bigg] \;,\label{eq:boltzmann2}
\end{align}
where $\eta_a \equiv n_a/n_\gamma$, $z\equiv M_N/T$, and $H_N$ is the Hubble rate at $T=M_N$. Here, $\gamma_D$ is the reaction density for the decays $N \leftrightarrow LH$, $H \leftrightarrow NL$, including thermal corrections, whose expression can be found in Ref.~\cite{Hambye:2016sby}; $\gamma_{NN}$ denotes collectively the one for the processes $NN \leftrightarrow f\bar f \,,\,  Z' Z'\,,\, Z' S\,,\, SS$, which can be calculated from the cross sections given in App.~\ref{sec:formulae}, the factor of 2 in \eqref{eq:boltzmann} being due to the fact that the two right-handed neutrinos are involved in the process; finally, the remaining reaction densities for $\Delta L=1$ scatterings can be found in Ref.~\cite{Giudice:2003jh}. We start the evolution from a zero lepton asymmetry and $N$ at equilibrium, since for the values of $g'$ considered in this work it is safe to assume that equilibration has occurred, possibly above the $B-L$ phase transition. In Eq.~\eqref{eq:boltzmann2}, $\epsilon_{\rm CP}(z)$ denotes the $\rm CP$ asymmetry, which we will assume to take a constant value $\epsilon$ (e.g.~its maximal value $\epsilon=1$), when either of the $\mathrm{CP}$-violating processes $N \leftrightarrow LH$, $H \leftrightarrow NL$ is kinematically allowed, after taking into account thermal effects~\cite{Hambye:2016sby}, and zero otherwise. However, notice that for low $M_N$ in the \unit{GeV} range, thermal effects imply that  maximal values for the asymmetry $\epsilon_{\rm CP} \sim 1$ cannot be realized~\cite{Hambye:2016sby}, and hence the bounds given below should be seen as very conservative in this regime.

Notice that the interactions beyond the minimal seesaw model, i.e.~the ones involving $Z'$ and $S$, do not enter the equation for the asymmetry~\eqref{eq:boltzmann2} at this order and therefore cannot wash out the asymmetry. However, they have a major role in keeping $N$ close to equilibrium, see~\eqref{eq:boltzmann}, and therefore can heavily dilute the generated asymmetry.

The Boltzmann equations~\eqref{eq:boltzmann} and~\eqref{eq:boltzmann2} do not take into account flavour effects in the charged-lepton sector. However, these are not expected to have a major impact on the results~\cite{Hambye:2016sby}, since here leptogenesis  typically occurs in the strong-washout regime $\tilde{m} \equiv v^2 (y y^\dagger)/M_N \gg \unit[2]{meV}$. Quantum coherences in the charged-lepton sector can change significantly the lepton asymmetry for $M_N \sim \unit[200]{GeV}$~\cite{Dev:2015vra} if the Yukawa couplings are very large, namely $y \sim 10^{-3}$, which can be  realized naturally in models with additional symmetries. However, since the asymmetry typically will be maximal for lower values of $y$, we may safely neglect this effect. We have also conservatively neglected the effect of tracking the evolution of the different $N$'s, since this can only slightly increase the washout of the asymmetry~\cite{Frere:2008ct}, unless one is interested in very large values of $y$ in models with additional leptonic symmetries. Again, as discussed above, we have safely neglected this. Finally, quantum coherences in the heavy-neutrino sector can give an additional $\mathcal{O}(1)$ contribution to the asymmetry~\cite{Dev:2015vra,Dev:2014wsa}, but this effect will not have a significant impact on the results below and thus it has been neglected, for simplicity, in Eqs.~\eqref{eq:boltzmann} and~\eqref{eq:boltzmann2}.

For $M_N$ in the \unit{GeV} range, in addition to thermal leptogenesis via $N$ and $H$ decay, as described by~\eqref{eq:boltzmann} and~\eqref{eq:boltzmann2}, an asymmetry can be generated via the Akhmedov--Rubakov--Smirnov (ARS) mechanism~\cite{Akhmedov:1998qx,Asaka:2005pn,Canetti:2012kh}. However, this effect will be highly suppressed by the fact that the evolution starts from a state of thermal equilibrium, as discussed above. Moreover, since the bounds below will be basically due to the equilibration of $N$ due to gauge interactions, which occurs also for the ARS mechanism for the same values of $g'$, the inclusion of this mechanism (for which it is difficult to perform general analyses; see e.g.~Refs.~\cite{Hernandez:2015wna,Drewes:2016gmt,Hernandez:2016kel}) would not significantly change our results.

\begin{figure}[t]
\includegraphics[width=0.45\textwidth]{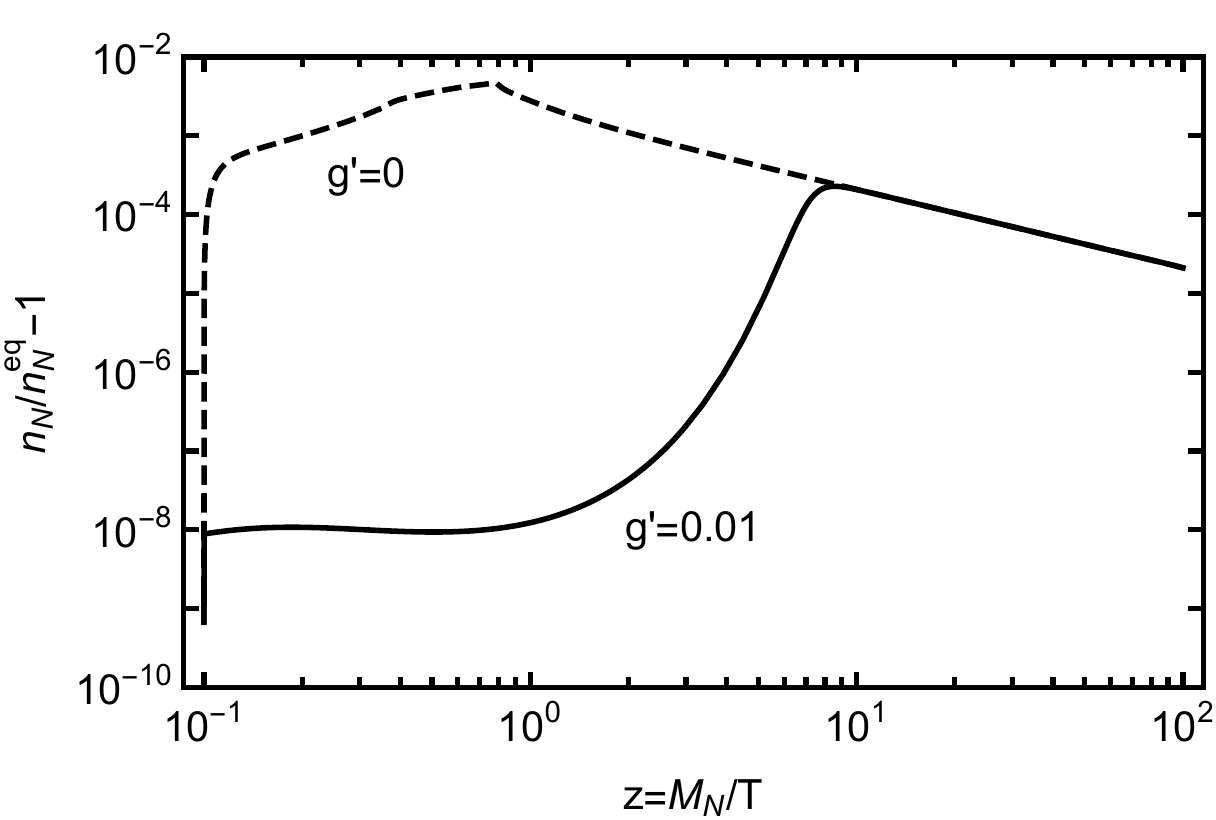}
\caption{
The deviation of the $N$ number density from equilibrium in the presence (continuous line) or absence (dashed line) of gauge interactions. For illustrative purposes, we have chosen $M_N=\unit[10]{TeV}$, $M_{Z'} = \unit[5]{TeV}$, $M_S=\unit[30]{TeV}$, and $\tilde{m} = \unit[1]{eV}$.
}
\label{fig:N_evolution}
\end{figure}

In Fig.~\ref{fig:N_evolution} we show the evolution of the $N$ number density as a function of the temperature, for a typical choice of the model parameters, in the presence or absence of gauge interactions. Although in low-scale leptogenesis the right-handed neutrino number density typically does not depart much from thermal equilibrium, the presence of gauge interactions can suppress further the departure from equilibrium by many orders of magnitude (until their decoupling), thus diluting strongly the asymmetry, as explained above.

\begin{figure}[b]
\includegraphics[width=0.45\textwidth]{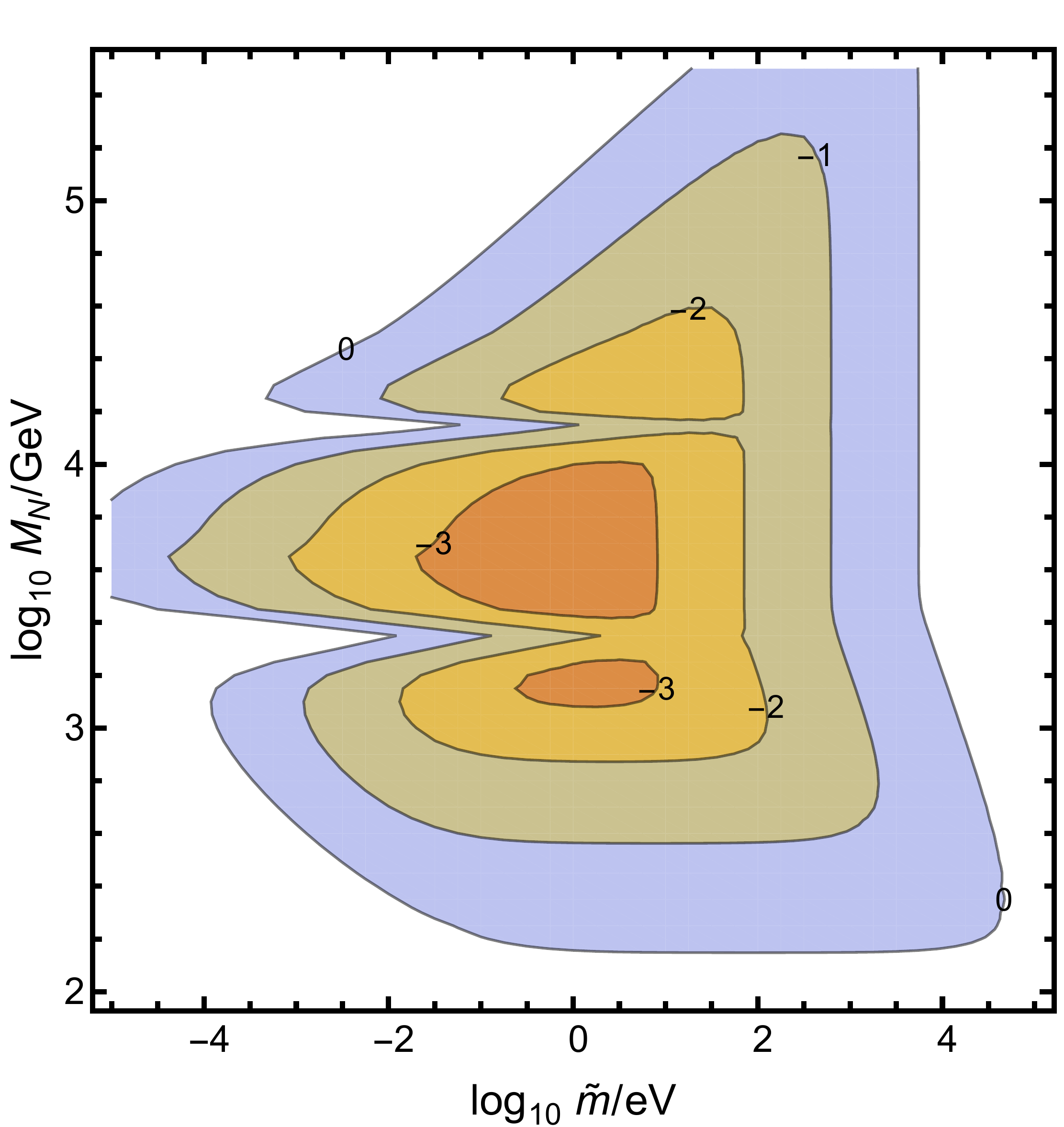}
\caption{
Contours of $\log_{10}\epsilon$ needed for successful leptogenesis for $M_{Z'} = \unit[5]{TeV}$, $M_S = \unit[30]{TeV}$, and $g' = 0.2$.
Perturbative unitarity excludes the region $M_N >\unit[44]{TeV}$.
}
\label{fig:mtilde}
\end{figure}

In Fig.~\ref{fig:mtilde} we show an example for the required CP asymmetry $\epsilon$ for a given set of parameters as a function of $\tilde{m}$. Clearly visible are the dips at the resonances $M_N \sim M_{Z'}/2$ and $M_N\sim M_S/2$, where the rates $NN\leftrightarrow \bar f f$ and $NN\leftrightarrow Z'Z'$ become resonantly enhanced and make leptogenesis more difficult, especially in the natural-seesaw regime $\tilde{m} \sim \unit[50]{meV}$. For very large $\tilde{m}$ leptogenesis becomes difficult because of the strong washout of the generated asymmetry; for small $\tilde{m} \lesssim \unit[10^{-4}]{eV}$ the generation of the asymmetry is suppressed by the slow rate of the CP-violating decays.
In the following we will always maximize the final asymmetry with respect to $\tilde{m}$ in order to get conservative limits.

\subsection{Decoupled \texorpdfstring{$S$}{S}} 

\begin{figure}[t]
\includegraphics[width=0.48\textwidth]{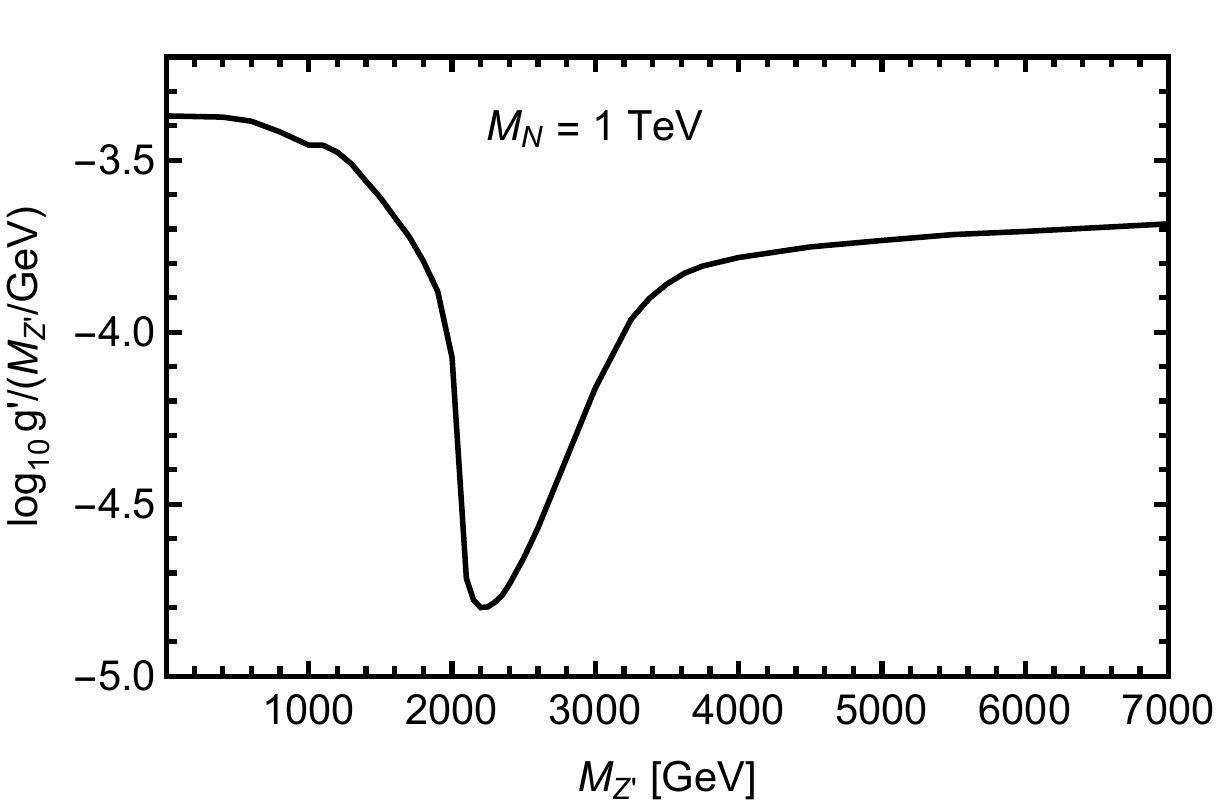}
\caption{
Upper limit on $g'/M_{Z'} = (2 w)^{-1}$ vs.~$M_{Z'}$ for decoupled $S$, $M_N = \unit[1]{TeV}$, and $\epsilon=1$. 
}
\label{fig:func_of_mZp}
\end{figure}

Let us consider the decoupled-scalar limit $M_S \to \infty$ first: 
for the extreme case $M_{Z'} \gg M_N$, only the rate $NN\leftrightarrow f\bar f$ is relevant, which is proportional to $w^{-4}$. Limits are then cast on the VEV $w$ as 
\begin{align}
w/\unit[12]{TeV} \ > \ 10^{-M_N/\unit[1.5]{TeV}} \;,
\end{align} 
for $\epsilon=1$ in the region $\unit[200]{GeV} \lesssim M_N \lesssim \unit[1]{TeV}$, and stronger for lower $M_N$. If the $Z'$ is light enough (roughly $2 M_N < M_{Z'} \lesssim 3.5 M_N$) the rate $NN\leftrightarrow f\bar f$ is resonantly enhanced (see Fig.~\ref{fig:func_of_mZp}), which will severely increase the limit on $w$. For $M_{Z'} < 2M_N$, this $s$-channel resonance disappears, and for $M_{Z'} \lesssim M_N$ the channel $NN\leftrightarrow Z'Z'$ opens up.\footnote{The threshold for $NN \to Z'Z'$ is technically not $M_N$ but the relevant temperature $T \gtrsim 2 M_{Z'}$ and analogously for the $N N \to S S$ process.\label{temperaturethreshold}} Since the rate $NN\leftrightarrow Z'Z'$ grows for small $M_{Z'}$ like $(g'/M_{Z'})^4$ due to the dominant emission of the would-be Goldstone boson,\footnote{In the limit $g'\to 0$ the $U(1)_{B-L}$ becomes a \emph{global} symmetry and the longitudinal $Z'$ component $\mathrm{Im}(\Phi)$ becomes the (massless) Majoron, with qualitatively similar impact on leptogenesis~\cite{Pilaftsis:2008qt,Gu:2009hn,Sierra:2014sta}.} limits on the $B-L$ parameter space for $M_{Z'} \ll M_N$ depend again on the VEV,
\begin{align}
w/\unit[5]{TeV} \ > \ 10^{-M_N/\unit[1.5]{TeV}}\;,
\end{align}
if we assume a maximal CP asymmetry $\epsilon =1$, up to the perturbative-unitarity bound $M_N < \unit[1.6]{TeV}$ (see discussion below).

\begin{figure}[t]
\includegraphics[width=0.46\textwidth]{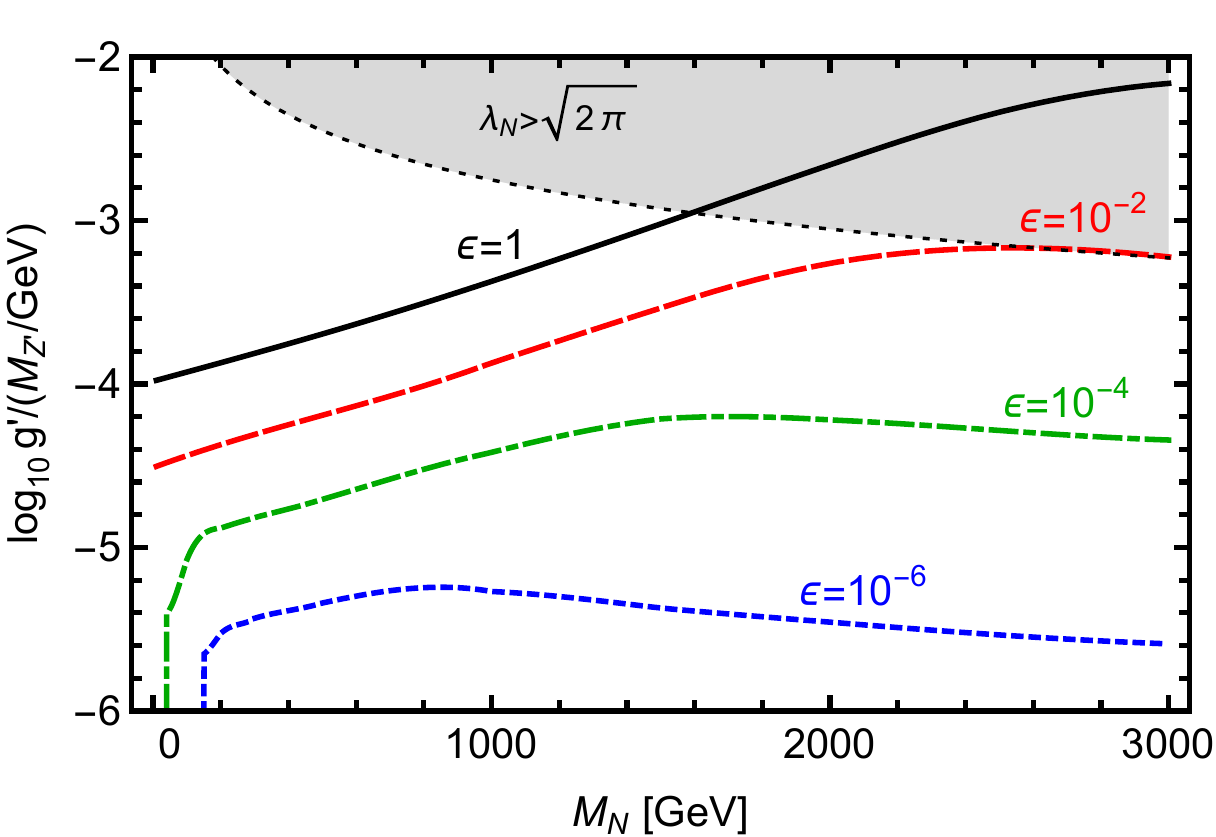}
\caption{
Upper limit on $g'/M_{Z'}$ vs.~$M_N$ for $M_{Z'} = \unit[1]{GeV}$ (valid for any $M_{Z'}\ll M_N$) and decoupled $S$ for various CP asymmetries. In the shaded region perturbative unitarity would be violated.
}
\label{fig:varying_eps}
\end{figure}

Let us remind the reader that it is difficult to actually obtain CP asymmetries $\epsilon$ of order 1, since this requires two of the Majorana neutrinos to be highly degenerate in order to resonantly enhance the asymmetry~\cite{Pilaftsis:1997jf},
\begin{align}
M_{N_i}-M_{N_j} \sim \Gamma_{i,j}/2\ll M_{N_{i,j}}\,,
\end{align}
where $\Gamma_{j}$ denotes the decay width of $N_{j}$.
In light of this, let us discuss the effect of smaller $\epsilon$.
As can be seen from Fig.~\ref{fig:varying_eps}, lower values of $\epsilon$ result in stronger limits on $w$ but are difficult to give in analytic form. In addition, we see the increasing lower bound on $M_N$~\cite{Hambye:2016sby}. Notice that, as discussed above, for $M_{Z'} \ll M_N $ the relevant rate depends on the combination $g'/M_{Z'}$; therefore, the results shown in Fig.~\ref{fig:varying_eps}, although explicitly given for $M_{Z'} = \unit[1]{GeV}$, are of general validity, as long as $Z'$ is much lighter than $N$.

\subsection{The effect of \texorpdfstring{$S$}{S}}

\begin{figure}[t]
\includegraphics[width=0.48\textwidth]{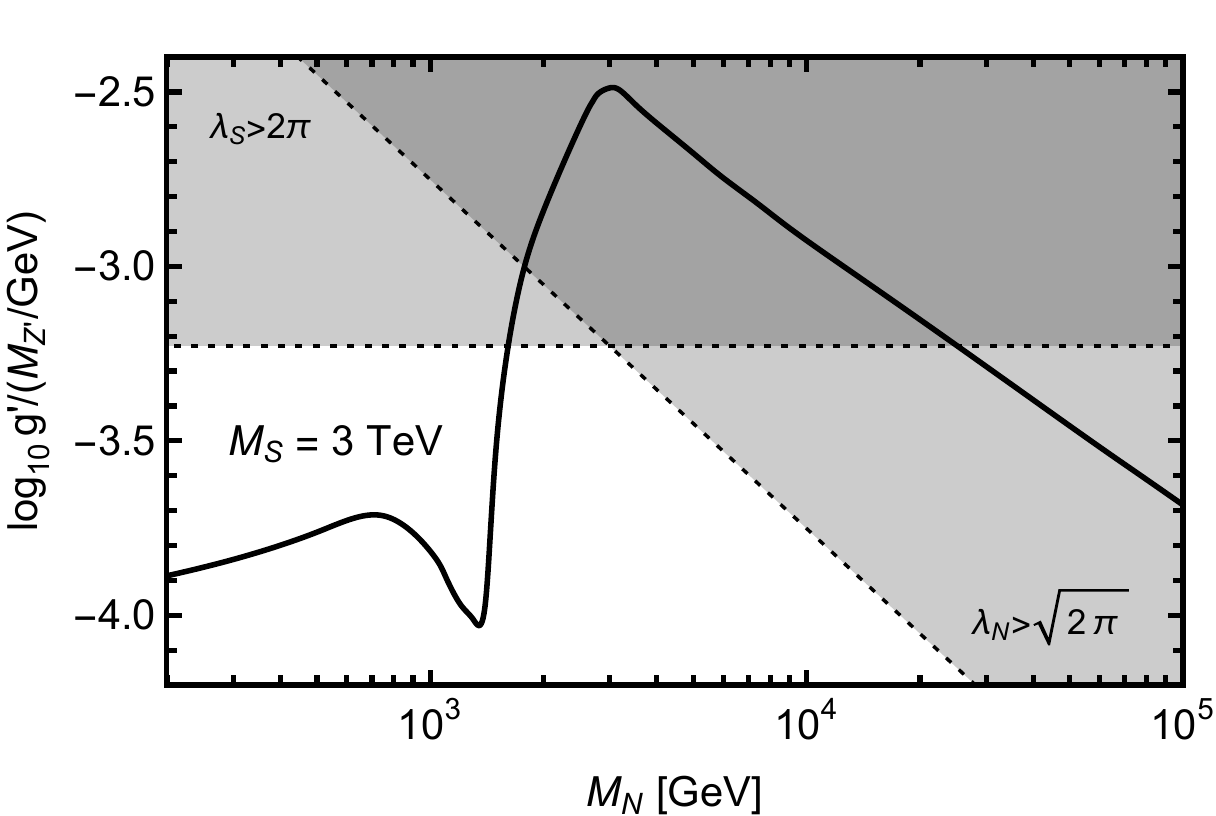}\\
\includegraphics[width=0.48\textwidth]{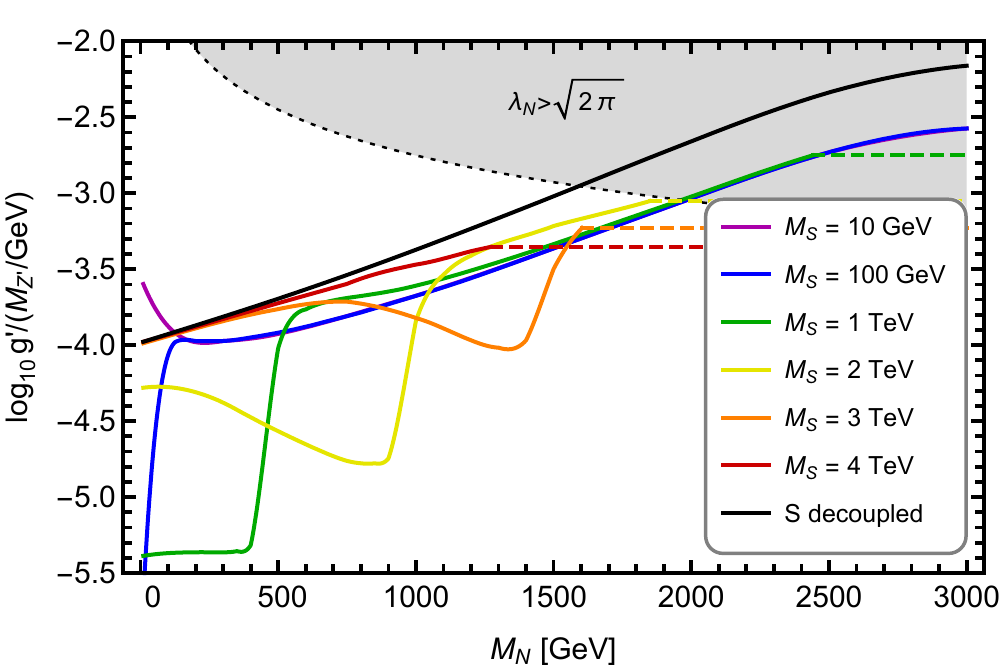}
\caption{
Upper limit on $g'/M_{Z'}$ vs.~$M_N$ for $M_{Z'} = \unit[1]{GeV}$ (valid for any $M_{Z'}\ll M_N$), $\epsilon=1$, and various $S$ masses. The shaded regions (top and bottom) and the dashed parts of the lines (bottom) correspond to the perturbative-unitarity bounds~\eqref{eq:pert}.
}
\label{fig:varying_mS}
\end{figure}

The previous results are fairly simple because we ignored/decoupled the additional new particle, the scalar $S$.
This is of course a much too simplifying assumption, as $S$ is required to unitarize e.g.~the $NN\to Z'Z'$ cross section. Differently stated, since the quartic interaction $S^4$ has a coupling constant $\lambda_S = M_S^2/2w^2$, a hierarchy $M_S\gg w$ leads to a non-perturbative coupling, outside of our region of calculability. A similar argument holds for the Yukawa coupling $\lambda_N$. 
We adopt the perturbative-unitarity limits from Refs.~\cite{Kahlhoefer:2015bea,Duerr:2016tmh}, which can be cast in the form 
\begin{align}\label{eq:pert}
M_{N,S} \leq \sqrt{4\pi} w \qquad \text{or} \qquad \lambda_N \leq \sqrt{2\pi}, \; \lambda_S \leq 2\pi \,.
\end{align}
Lowering $M_S$ to perturbative values will typically \emph{strengthen} the limits on the $B-L$ parameter space, because the destructive interference in the $NN\to Z'Z'$ rate is less relevant than the $s$-channel $S$ resonance that becomes readily accessible in the thermal bath. Lowering $M_S$ below $2 M_N$ turns this resonance off, but opens the channel $NN \to SZ'$ if the $Z'$ is sufficiently light. Finally, for $M_S \lesssim M_N$, the channel $NN\to SS$ opens up (see also Ref.~\cite{Iso:2010mv}).\footnoteref{temperaturethreshold}
All of this is illustrated in Fig.~\ref{fig:varying_mS}. We see that the bounds on $M_{Z'}/g'$ can increase by more than an order of magnitude compared to the decoupled $S$.
Only for $M_S \lesssim \unit[20]{GeV}$ do the bounds become \emph{weaker} than those of the decoupled $S$ case, and only for the similarly small $M_N$.

From Figs.~\ref{fig:func_of_mZp}, \ref{fig:varying_eps} and~\ref{fig:varying_mS} we see that the weakest, most conservative bound on the $B-L$ parameter space compatible with perturbative unitarity arises for the hierarchy $M_{Z'} \lesssim M_N\ll M_S$ and reads 
\begin{align}\label{eq:loose_bound}
M_{Z'}/g' \ = \ 2w \ \gtrsim \ \unit[1]{TeV}\,,
\end{align}
achievable at $M_N\sim \unit[1.6]{TeV}$.
For more realistically obtainable CP asymmetries $\epsilon = 10^{-2}$ ($10^{-4}$), the most conservative bound can be read off of Fig.~\ref{fig:varying_eps} as $w\gtrsim \unit[800]{GeV}$ ($\unit[8]{TeV}$).
We stress that these limits increase rather dramatically if the hierarchies $M_{Z'} \lesssim M_N\ll M_S$ are invalid.

\section{Phenomenology}
\label{sec:pheno}

\subsection{Light \texorpdfstring{$Z'$}{Z'}}

\begin{figure*}[t]
\includegraphics[width=0.8\textwidth]{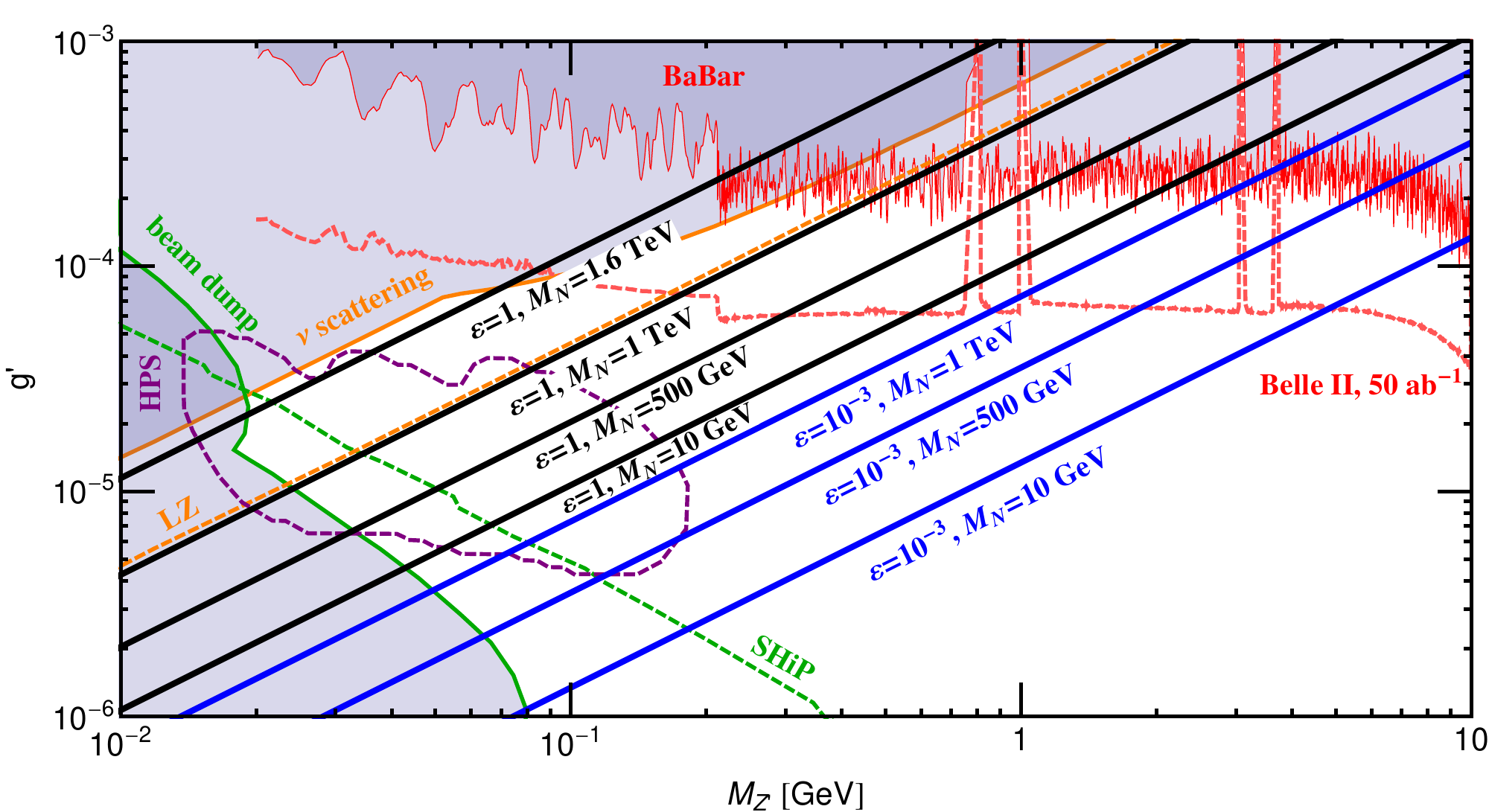}
\caption{
Parameter space of a gauge boson $Z'$ coupled to $B-L$. The shaded areas are excluded at $90\%$~C.L., and
future reach is shown in dashed lines; see text for details.
The diagonal black and blue lines show the upper bound for successful leptogenesis with CP asymmetries $\epsilon = 1$ and $10^{-3}$, respectively, for various $N$ masses.
The black line $\epsilon=1$, $M_N = \unit[1.6]{TeV}$ corresponds to the most conservative limit from leptogenesis within the perturbative region.
}
\label{fig:limits}
\end{figure*}

Having derived limits on the $B-L$ parameter space from successful leptogenesis, we can compare to existing bounds and regions to be probed in the near future, focusing for now on $\unit[10]{MeV}\leq M_{Z'} \leq \unit[10]{GeV}$. We follow Ref.~\cite{Heeck:2014zfa} to translate limits from beam-dump experiments~\cite{Andreas:2012mt}, BaBar~\cite{Lees:2014xha}, and $\nu_{e,\mu}$--$e^-$ scattering data~\cite{Harnik:2012ni,Bilmis:2015lja} (see Fig.~\ref{fig:limits}). Also shown is the potential reach of the proposed SHiP experiment~\cite{Alekhin:2015byh}, adopted from Refs.~\cite{Gorbunov:2014wqa,Kaneta:2016vkq}; of Belle~II (for an integrated luminosity of $\unit[50]{ab^{-1}}$)~\cite{Ferber:2015jzj,Hearty}; and of the Heavy Photon Search (HPS) experiment~\cite{Celentano:2014wya}, naively converted from limits on hidden photons.
Finally, solar neutrino scattering with electrons in future second-generation xenon-based dark matter direct detection experiments such as LZ can further probe the parameter space with limits up to $w\simeq \unit[1]{TeV}$ (LZ line in Fig.~\ref{fig:limits})~\cite{Cerdeno:2016sfi}.
Prospects for LHCb sensitivity require a more careful adaption from the hidden photon case~\cite{Ilten:2015hya,Ilten:2016tkc} and are beyond the scope of this article.

Notice that the observation of a $U(1)_{B-L}$ $Z'$ with $M_{Z'}/g' \lesssim\unit[7]{TeV}$ (much stronger for $M_{Z'} \lesssim\unit[5]{GeV}$~\cite{Heeck:2014zfa}) would already establish that neutrinos cannot be Dirac particles, because the $\nu_R$ would be thermalized by the $Z'$ interactions and disturb Big Bang nucleosynthesis by contributing to $N_\mathrm{eff}$. If neutrinos are Majorana particles, the same argument requires the sterile Majorana partners $N$ to be typically heavier than the MeV scale. As shown here, successful leptogenesis in the presence of such a $Z'$ severely increases this lower bound on $M_N$.

As shown above, the most conservative leptogenesis constraint~\eqref{eq:loose_bound} (obtained for $\epsilon=1$, $M_{Z'} \lesssim M_N\ll M_S$) already implies $M_{Z'}/g' \gtrsim \unit[1]{TeV}$, and is hence stronger than the neutrino scattering constraints below $M_{Z'} = \unit[0.1]{GeV}$ (Fig.~\ref{fig:limits}). For $M_{Z'} > \unit[0.1]{GeV}$, neutrino scattering limits, among others, are superior to the most conservative leptogenesis constraint.
This is no longer true for realistically achievable CP asymmetries $\epsilon \ll 1$, which can easily provide limits even beyond future neutrino-scattering reach. Since the largest realistic value for $\epsilon$ is however impossible to quantify (or to measure), this is at best a qualitative constraint on the parameter space.
Nevertheless, even for $\epsilon=1$ one can obtain strong limits if the hierarchy $M_{Z'} \lesssim M_N\ll M_S$ is invalid, which is experimentally testable by observing both $Z'$ and $N$, the scalar $S$ being practically impossible to observe unless we turn on scalar mixing. In particular, the requirement of successful leptogenesis becomes competitive to constraints from direct searches for $M_N\ll \unit{TeV}$, and it is precisely this region where one could hope to find $N$.

Since leptogenesis is infamously hard to verify as a mechanism, a more relevant alternative question is thus whether it is falsifiable~\cite{Frere:2008ct,Deppisch:2013jxa,Deppisch:2015yqa}. 
As an optimistic example, let us assume that a $Z'$ is found right around the corner of existing limits at $M_{Z'} \simeq \unit[20]{MeV}$ with $g' \sim 2\times 10^{-5}$, which implies a VEV $w\sim \unit[500]{GeV}$ and offers ample opportunity to study the new particle in different experiments. Successful leptogenesis can only be obtained for $M_N \gtrsim \unit[1.5]{TeV}$ (Fig.~\ref{fig:varying_eps}), so the observation of a \emph{sub-TeV} right-handed neutrino would falsify leptogenesis. Since $N$ masses up to $\mathcal{O}( \unit[100]{GeV})$ can be probed at FCC-ee, ILC and CEPC~\cite{Blondel:2014bra,Antusch:2015mia,Antusch:2016vyf} -- depending on the active--sterile mixing angle -- there indeed exists the possibility to falsify leptogenesis in models where the right-handed neutrino has additional couplings to a neutral gauge boson.
Here we are of course glossing over the intricacies of establishing the true identities of $Z'$ and $N$, i.e.~all their relevant couplings.

\subsection{Heavy \texorpdfstring{$Z'$}{Z'}}

\begin{figure}[t]
\includegraphics[width=0.48\textwidth]{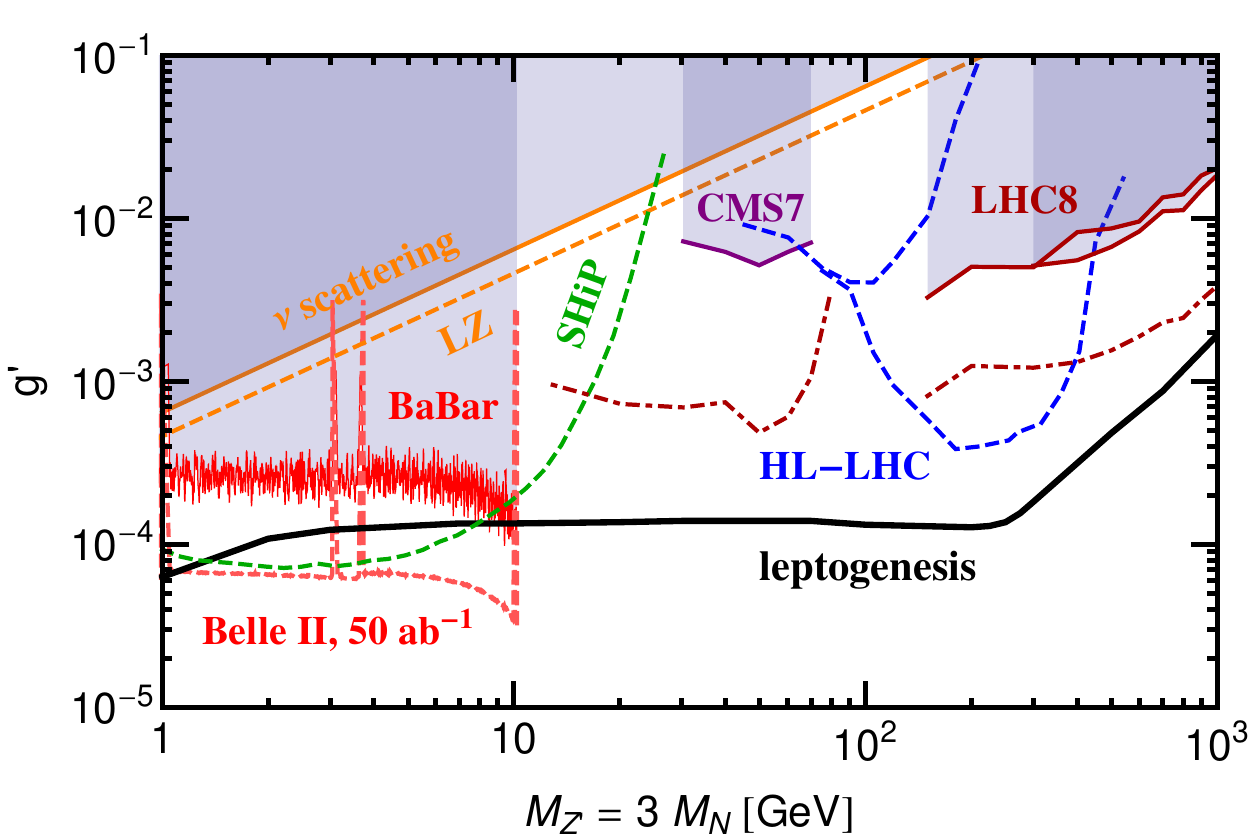}
\caption{
Upper limit on $g'$ vs.~$M_{Z'} = 3 M_N$ for decoupled $S$ and $\epsilon=1$ (thick black line).
Existing limits (shaded) and future prospects (dashed) under the assumption $M_{Z'}/M_N = 3$ and the seesaw relation for the active--sterile mixing angle are taken from Ref.~\cite{Batell:2016zod}. In particular, the dark red dot-dashed (blue dashed) lines show the projections for the HL-LHC ($\unit[3]{ab^{-1}}$) in the channel $pp\to Z'\to \ell^+\ell^-$ ($pp\to Z' \to NN$ with displaced vertices), while the dashed green line shows the SHiP reach for displaced vertices. For $M_N$ lighter than a few \unit{GeV} the leptogenesis bound reported should be considered very conservative; see the discussion in Sec.~\ref{sec:boltz}.
}
\label{fig:fixed_ratio}
\end{figure}

Let us discuss a second region of interest, with $M_{Z'} >\unit[10]{GeV}$, which in particular opens up the leptogenesis region around the $Z'$ resonance $M_{Z'} \sim 2 M_N$, since the lower bound on $M_N$ is around a few GeV~\cite{Hambye:2016sby}. 
On the experimental side, there are no $B$-factory limits for $M_{Z'} >\unit[10]{GeV}$, so the strongest constraints (and future prospects~\cite{Cerdeno:2016sfi}) come from neutrino scattering, specifically CHARM-II~\cite{Vilain:1994qy,Bilmis:2015lja}, together with LHC searches~\cite{Hoenig:2014dsa} (see also Refs.~\cite{Alves:2015mua,Klasen:2016qux} for dilepton bounds); see Fig.~\ref{fig:fixed_ratio}.
As has been emphasized in Ref.~\cite{Batell:2016zod} (see also Refs.~\cite{Huitu:2008gf,Blanchet:2009bu}), the additional gauge interactions of $N$ can severely improve the discovery prospects of both $N$ and $Z'$, provided the decay of $N$ is slow enough to lead to displaced vertices. This is naturally the case if the active--sterile mixing angle $\theta$ fulfills the naive seesaw relation $\theta^2 \sim M_\nu/M_N$, which implies $\tilde{m}\sim \unit[50]{meV}$. The second ingredient for the displaced-vertex observation of $N$ is the hierarchy $M_{Z'} > 2M_N$, resulting in a very efficient on-shell production of $Z'$ at the LHC (or SHiP), followed by $Z'\to NN$. Due to the very low background, these displaced-vertex searches could in the future be even more sensitive to a $Z'$ than the standard dilepton channel $pp\to Z' \to \ell\ell$~\cite{Batell:2016zod}.

\begin{figure}[t]
\includegraphics[width=0.48\textwidth]{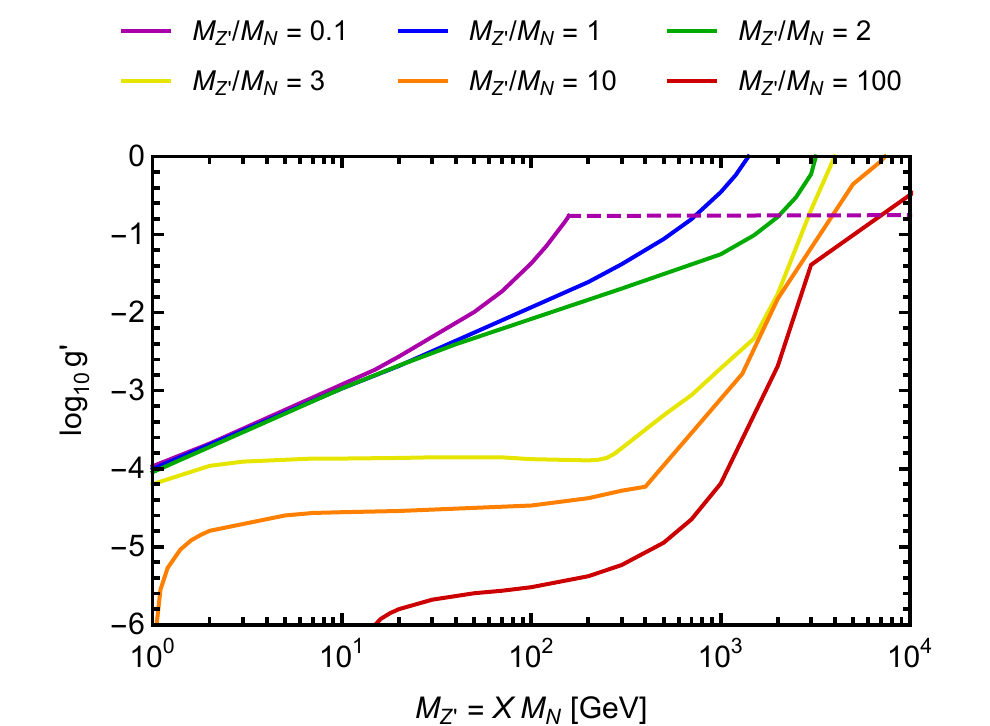}
\caption{
Upper limit on $g'$ vs.~$M_{Z'} = X M_N$ for $X\in \{0.1,1,2,3,10,100\}$, decoupled $S$ and $\epsilon=1$.
The dashed part of the line corresponds to the perturbative-unitarity bound~\eqref{eq:pert}. For $M_N$ lighter than a few \unit{GeV} the leptogenesis bounds reported should be considered very conservative; see the discussion in Sec.~\ref{sec:boltz}.
}
\label{fig:fixed_ratio_X}
\end{figure}

For leptogenesis, the hierarchy $M_{Z'} > 2 M_N$ implies that the limits on $w$ become stronger than before because the thermal distribution of $N$ energies in the early Universe makes it very easy to hit the $s$-channel $Z'$ resonance, if the mass of $Z'$ is below the \unit{TeV} range.
Taking as a benchmark the fixed ratio $M_{Z'}/M_N = 3$ (as in Ref.~\cite{Batell:2016zod}), we can give the conservative leptogenesis constraints for $\epsilon=1$ (decoupled $S$); see Fig.~\ref{fig:fixed_ratio}. (Our results agree to a good accuracy with the simplified analysis of Ref.~\cite{Blanchet:2009bu}.)
Some comments are in order: the future prospects for observing $Z'$ and $N$ in Fig.~\ref{fig:fixed_ratio} assume the production of \emph{one} pair of right-handed neutrinos with decay length fixed by the naive seesaw relation $\theta^2 \sim M_\nu/M_N$ for the mixing angle. The limits and prospects should thus slightly increase for our case with (at least two) \emph{degenerate} $N$. Furthermore, fixing $\theta$ implies a fixed $\tilde{m}$, a parameter we varied to maximize $\epsilon$ in our leptogenesis limits, so our limits will also increase (slightly).
Increasing the ratio $M_{Z'}/M_N$ far above 3 will not change much the current or projected limits of the dilepton channel, which rescale slightly with $\mathrm{BR}(Z'\to \ell\ell)$, but will reduce the efficiency for reconstructing the displaced vertex of the boosted $N$. The leptogenesis constraints on the other hand become stronger still for $M_{Z'}/M_N \gg 3$ (Fig.~\ref{fig:fixed_ratio_X}), making it possible to exclude leptogenesis robustly through a double observation of $Z'$ and $N$ with $M_{Z'}/M_N \gtrsim 3$. 
Let us emphasize again that the $\epsilon=1$ limit given in Figs.~\ref{fig:fixed_ratio} and~\ref{fig:fixed_ratio_X} for $M_N$ lighter than a few GeV is extremely conservative (see Sec.~\ref{sec:boltz}) and will realistically fall below even the Belle-II reach in Fig.~\ref{fig:fixed_ratio}.

\section{Conclusion}
\label{sec:conclusion}

The matter--antimatter asymmetry of our Universe is a longstanding mystery given our confidence in the (inflationary) Big Bang theory. Leptogenesis is championed as a simple explanation that also resolves the neutrino mass problem of the SM.

In this paper we have performed a comprehensive analysis of leptogenesis in the presence of a neutral gauge boson $Z'$ interacting with the right-handed neutrinos $N$, taking the gauged $U(1)_{B-L}$ model as a benchmark scenario. The gauge interactions put $N$ in thermal equilibrium at high temperatures; the additional annihilation channel $NN\to \bar{f}f$ via heavy $s$-channel $Z'$ or $NN\to Z'Z'$ for light $Z'$ will dilute the lepton asymmetry, allowing us to provide a very conservative lower limit $M_{Z'}/g' \gtrsim \unit[1]{TeV}$ for successful leptogenesis. This limit increases drastically for realistically achievable CP asymmetries or if the hierarchy $M_{Z'} \lesssim M_N$ is violated. 
The latter case, with $M_{Z'}>2M_N$, is of particular interest because it allows for an efficient production of $N$ via on-shell $Z'\to NN$, potentially followed by a displaced-vertex decay of $N$. This is a promising detection channel for both $Z'$ and $N$ at SHiP or the (HL-)LHC. As it can be seen from Fig.~\ref{fig:fixed_ratio}, leptogenesis provides extremely strong constraints in this region of parameter space, so that any observation of $Z'$ and $N$ via displaced vertices would effectively rule out leptogenesis.

We have also considered for the first time the opposite hierarchy, $M_{Z'} < 2M_N$, and shown that it still poses strong limits on the $Z'$ parameter space (Fig.~\ref{fig:limits}). Even here it is possible to falsify leptogenesis by detecting both $Z'$ and $N$ in a large region of parameter space. In this case of a light $Z'$, whose discovery prospects will increase significantly in the near future, it has been important to consider $NN \to Z'Z'$ processes, in addition to $NN \to f \bar{f}$, which is in turn dominant for heavy $Z'$. A careful treatment of the scalar $S$, responsible for the breaking of $B-L$, has shown that its effect can make the bounds on successful leptogenesis even more than an order of magnitude stronger.

\section*{Acknowledgements}
JH thanks Sebastian Ohmer, Felix Kahlhoefer, and Camilo Garcia-Cely for discussions and Brian Shuve and Christopher Hearty for kindly sharing their data shown in Fig.~\ref{fig:fixed_ratio}.
We thank Jean-Marie Fr\`ere and Thomas Hambye for comments on the manuscript.
DT is supported by the Belgian Federal Science Policy through the Interuniversity Attraction Pole P7/37.
JH is a postdoctoral researcher of the F.R.S.-FNRS.
We acknowledge the use of \texttt{Package-X}~\cite{Patel:2015tea}.

\appendix

\section{Formulas}
\label{sec:formulae}

For the convenience of the reader, we collect the relevant decay widths of the new particles $Z'$ and $S$ into leptons $\ell \in \{e,\mu,\tau\}$; quarks $q\in \{u,d,c,s,t,b\}$; light neutrinos $\nu \in \{\nu_e,\nu_\mu,\nu_\tau\}$; and heavy neutrinos $N\in \{N_1,N_2,N_3\}$, using $\alpha'\equiv {g'}^2/4\pi$:	
\begin{align}
\Gamma (Z' \to \bar{\ell} \ell) &= \frac{\alpha'  M_{Z'}}{3} \left(1 + \frac{2 M_\ell^2}{M_{Z'}^2}\right) \sqrt{1 -  \frac{4 M_\ell^2}{M_{Z'}^2}}\,,\\
\Gamma (Z' \to \bar{q} q) &= \frac{\alpha'  M_{Z'}}{9} \left(1 + \frac{2 M_q^2}{M_{Z'}^2}\right) \sqrt{1 -  \frac{4 M_q^2}{M_{Z'}^2}}\,,\\
\Gamma (Z' \to N N ) &= \frac{\alpha'  M_{Z'}}{6}  \left( 1 -  \frac{4 M_N^2}{M_{Z'}^2} \right)^{3/2} ,\\
\Gamma (Z' \to \nu \nu) &= \frac{\alpha'  M_{Z'}}{6} \,,\\
\Gamma (S \to N N) &= \alpha'  M_S \frac{M_N^2}{M_{Z'}^2}\left(1 -  \frac{4 M_N^2}{M_S^2}\right)^{3/2} ,\\
\begin{split}
\Gamma (S \to Z' Z') &= \alpha'  M_S \frac{ M_S^2}{2 M_{Z'}^2}\sqrt{1 -  \frac{4 M_{Z'}^2}{M_S^2}}\\
&\quad \times  \left(1 - \frac{4 M_{Z'}^2}{M_S^2} +\frac{12 M_{Z'}^4}{M_S^4} \right) .
\end{split}
\end{align}
The spin-averaged total cross section for $NN\to \bar f f$ via $s$-channel $Z'$ depends on the $B-L$ charge $Q_{B-L}$ and color multiplicity $N_c$ of the fermion $f$:
\begin{align}
\begin{split}
\sigma (NN&\to \bar f f) = \frac{N_c(f) {Q_{B-L}(f)}^2{g'}^4   }{12 \pi s  \left[\left(M_{Z'}^2-s\right)^2 + M_{Z'}^2 \Gamma_{Z'}^2\right]} \\
&\times\left(s+2 M_f^2\right)\sqrt{s-4 M_f^2} \sqrt{s-4 M_N^2} \,.
\end{split}
\end{align}
The spin-averaged cross section for $NN\to SS$ can be written as
\begin{widetext}
\begin{align}
\begin{split}
\sigma &= \frac{{g'}^4 M_N^2}{4 \pi M_{Z'}^4  \left(M_S^2-s\right)^2 \left(2 M_S^2-s\right) s \left(4 M_N^2-s\right)} \left[\frac{\left(s-2 M_S^2\right) \sqrt{\left(s-4 M_N^2\right) \left(s-4 M_S^2\right)}}{M_S^4+M_N^2 \left(s-4 M_S^2\right)} \right.\\
&\times\left\{-32 M_N^6 \left(M_S^2-s\right)^2+9 M_S^8 s+3 M_N^2 M_S^4 \left(s^2-6 M_S^4-16 M_S^2 s\right)+4 M_N^4 \left(20 M_S^6+4 M_S^4 s+4 M_S^2 s^2-s^3\right)\right\}\\
&+2 M_N^2 \left(M_S^2-s\right) \left\{18 M_S^6+32 M_N^4 \left(M_S^2-s\right)+10 M_S^4 s-11 M_S^2 s^2+s^3+16 M_N^2 \left(s^2-5 M_S^4+M_S^2 s\right)\right\} \\
&\left. \times\text{arctanh}\left[\frac{\sqrt{\left(s-4 M_N^2\right) \left(s-4 M_S^2\right)} \left(s-2 M_S^2\right)}{2 M_N^2 \left(s-4 M_S^2\right)-2 M_S^4+4 M_S^2 s-s^2}\right]
\right] ,
\end{split}
\end{align}
while the spin-averaged cross section for $NN\to Z'Z'$ (including $s$-channel $S$ exchange) is more complicated:
\begin{align}
\begin{split}
\sigma &= \frac{{g'}^4}{4 \pi M_{Z'}^8  \left[\left(M_S^2-s\right)^2 + M_S^2 \Gamma_S^2\right] \left(2 M_{Z'}^2-s\right) s \left(s-4 M_N^2\right)}
\left[ \frac{M_{Z'}^4 \left(2 M_{Z'}^2-s\right) \sqrt{\left(s-4 M_N^2\right) \left(s-4 M_{Z'}^2\right)}}{2 \left(M_{Z'}^4+M_N^2 \left(s-4 M_{Z'}^2\right)\right)} \right.\\
&\times \left\{-2 M_{Z'}^4 \left(M_{Z'}^2-4 M_N^2\right) \left(2 M_S^4 M_N^2+\left(M_S^4-8 M_S^2 M_N^2+48 M_N^4\right) M_{Z'}^2\right) \right.\\
&+M_{Z'}^2 \left(M_S^4 M_N^2 \left(-16 M_N^2+M_{Z'}^2\right)+4 M_S^2 \left(-2 M_N^2 M_{Z'}+M_{Z'}^3\right)^2+8 M_N^2 M_{Z'}^2 \left(-28 M_N^4+M_{Z'}^4\right)\right) s\\
&+ 2 \left(M_S^4 M_N^4+4 M_N^4 \left(M_S^2+8 M_N^2\right) M_{Z'}^2+M_N^2 \left(M_S^2+8 M_N^2\right) M_{Z'}^4+2 M_N^2 M_{Z'}^6-M_{Z'}^8\right) s^2\\
&\left. -M_N^2 \left(8 M_N^4+8 M_N^2 M_{Z'}^2+M_{Z'}^4\right) s^3\right\}
+\tfrac{1}{4} M_{Z'}^4 \left(M_S^2-s\right) \left\{-4 \left(M_S^2-8 M_N^2\right) M_{Z'}^6 \left(4 M_N^2-M_{Z'}^2\right) \right.\\
&-4 M_{Z'}^2 \left(32 M_N^4 M_{Z'}^2-8 M_N^2 M_{Z'}^4+M_{Z'}^6+M_S^2 \left(-4 M_N^4+3 M_N^2 M_{Z'}^2\right)\right) s\\
&+\left(-4 M_S^2 M_N^4+4 M_N^2 \left(M_S^2+8 M_N^2\right) M_{Z'}^2+\left(M_S^2+12 M_N^2\right) M_{Z'}^4\right) s^2\\
&\left.\left. -\left(2 M_N^2+M_{Z'}^2\right)^2 s^3\right\} \log\left[\left(\frac{2 M_{Z'}^2-s+\sqrt{\left(s-4 M_N^2\right) \left(s-4 M_{Z'}^2\right)}}{2 M_{Z'}^2-s-\sqrt{\left(s-4 M_N^2\right) \left(s-4 M_{Z'}^2\right)}}\right)^2\right]
\right] .
\end{split}
\end{align}
The expression for $NN\to SZ'$ is even more involved and will not be shown here. In the non-relativistic limit, all four annihilation cross sections $NN\to \bar ff$, $SS$, $Z'Z'$, $SZ'$ match those of Ref.~\cite{Duerr:2016tmh}. Notice that to calculate the relevant thermally averaged rates entering the Boltzmann equations, one needs the spin-\emph{summed} \emph{reduced} cross sections~\cite{Luty:1992un} which can be readily inferred from the formulas above.
\end{widetext}

\bibliographystyle{utcaps_mod}
\bibliography{BIB}

\end{document}